\documentclass[twocolumn,prb,showpacs]{revtex4b4}
\usepackage{graphicx}
\usepackage{dcolumn}
\usepackage{amsmath}
\begin{document}
\title{Quantum Monte Carlo treatment of elastic exciton-exciton scattering}
\author{J.~Shumway}
\email{jshumway@nrel.gov}
\altaffiliation{Present address: National Renewable Energy Laboratory,
Golden, CO 80401}
\author{D.~M.~Ceperley}
\affiliation{Department of Physics
and the National Center for Supercomputing Applications\\
University of Illinois at Urbana-Champaign,
1110 West Green Street,  Urbana, Illinois 61801}
\date{\today}
\begin{abstract}
We calculate cross sections for low energy elastic exciton-exciton
scattering within the effective mass approximation.
Unlike previous theoretical approaches, we give a complete, 
non-perturbative treatment of the four-particle scattering 
problem.  Diffusion Monte Carlo is used to calculate the
essentially exact
energies of scattering states, from which phase shifts are determined.
For the case of equal-mass electrons and holes, which is equivalent to 
positronium-positronium scattering, we find $a_s = 2.1 a_x$ 
for scattering of singlet-excitons and $a_s= 1.5 a_x$ for triplet-excitons,
where $a_x$ is the excitonic radius.
The spin dependence of the cross sections arises
from the spatial exchange symmetry of the scattering wavefunctions.
A significant triplet-triplet to singlet-singlet scattering process
is found, which is similar to reported effects in recent 
experiments and theory for excitons in quantum wells.  We also show
that the scattering length can change sign and diverge for some 
values of the mass ratio $m_h$/$m_e$, an effect not seen
in previous perturbative treatments.
\end{abstract}

%Intrinsic properties of excitons; optical absorption
%keywords: exciton, scattering, positronium, quantum monte carlo

\maketitle

\section{Introduction}
Excitons in semiconductors have been the subject of many
experimental and theoretical investigations of Bose condensation.
Low-energy exciton-exciton interactions are characterized by
the exciton-exciton scattering length, $a_s$, which
determines the thermodynamics of a low density
gas and is crucial for modeling the thermalization time
of a dilute exciton gas.  Despite its importance, the
exciton-exciton  scattering length is an elusive quantity, 
being difficult to measure experimentally or 
to estimate theoretically.

As is well-known in atomic physics, scattering lengths
can be extremely sensitive to the details of the interactions
between particles.  In particular, the existence of a weakly
bound or nearly bound state causes the scatter length to
become quite large.  Therefore, {\em a priori} one should
suspect that exciton-exciton scattering may be a very material
dependent property of semiconductors.  Reliable theoretical predictions
of exciton-exciton scatting lengths require both a very accurate
Hamiltonian for the semiconductor, and an accurate solution to the 
(four-particle) scattering problem. In this paper we 
provide an essentially exact solution to exciton-exciton 
scattering for a commonly used single-band effective mass Hamiltonian.
This solution allows us to study three important questions:
(1) how sensitive is the scattering length to the mass ratio $m_e/m_h$,
(2) how does the scattering length depend on spin states 
(singlet or triplet) of the scattering excitons,
and (3) to what degree can inter-exciton exchange of electrons or holes
cause excitons to scatter into different spin states?
This calculations also serve as a benchmark for the single
band limit of more complicated scattering Hamiltonians.

One experimental method for measuring the exciton scattering
cross section is to look at line width broadening of the 
recombination spectra in a gas of excitons.
Collisions between excitons increase the line width, causing
the line width to depend on the exciton-exciton scattering
rate, $n\sigma v$, where $n$ is the density and $v$ is a typical
exciton velocity.  Extracting cross sections from a
line width requires that (1) the density and velocity distribution
are known, and (2) elastic scattering is the fastest process.
As discussed below, $\mathrm{Cu_2O}$ is a good material for comparison
to the model studied in this work.  
Snoke {\em et al.}\cite{snoke90,snoke91} have performed such
experiments on $\mathrm{Cu_2O}$ and have found a line 
width broadening that suggests an upper bound of 
$4 a_x$ on the scattering length.  Although our simulations
do not exactly model $\mathrm{Cu_2O}$, we will compare our
results to this value.

\section{Theoretical background}
Theoretical approaches to this problem start with 
the effective mass approximation, in which the system under consideration
consists of two electrons, labeled $1$ and $2$, and two holes,
labeled $a$ and $b$.  The Hamiltonian is
\begin{eqnarray}
H&=&-\lambda_1\nabla_1^2 -\lambda_2\nabla_2^2
  -\lambda_a\nabla_a^2 -\lambda_b\nabla_b^2\nonumber\\
&&- r_{a1}^{-1} - r_{b2}^{-1}
  - r_{a2}^{-1} - r_{b1}^{-1}
  + r_{12}^{-1} + r_{ab}^{-1},
\label{hamil}
\end{eqnarray}
where $\lambda=\hbar^2/2m$.
The Hamiltonian has symmetry under exchange 
of electrons and exchange of holes, so eigenstates may be denoted by
two exchange quantum numbers.  The s-wave states
are symmetric under exchange of excitons; a condition
which is satisfied by states $\phi^{++}$ and $\phi^{--}$,
where the $+(-)$ signs refer to (anti)symmetry under
exchange of electrons and holes, respectively.
Although this Hamiltonian is a well-accepted model for
exciton-exciton scattering, we should point out a few of 
its deficiencies.  For small excitons, such as those in
$\mathrm{Cu_2O}$, that have radii not much larger than the
lattice spacing, non-parabolic terms in the kinetic energy
and other corrections to the potential energy may be necessary.
For many semiconductors, such as Si and Ge, the valence band
is a mixture of three bands and cannot be described by a single
parabolic band.  In the case of $\mathrm{Cu_2O}$, the valence
band is the parabolic spin-orbit split off band, and there are
fewer complications.  Interband exchange (virtual electron-hole
recombination) is an important effect that has been neglected, 
and could be modeled by an additional spin-dependent 
potential term.

This Hamiltonian also describes a family of scattering processes
for other particles, including hydrogen-hydrogen, positronium-positronium,
and muonium-muonium scattering.  The equal mass case is at an
extreme (positronium scattering), where the Born-Oppenheimer approximation
is the least applicable.

There have been several theoretical estimates of exciton-exciton
scattering for bulk systems\cite{haug76,elkomoss81,elkomoss84} and
quantum wells\cite{cuiti98,koh97} as well as calculations on
biexciton-biexciton scattering,\cite{cam97}
Only the bulk, elastic scattering calculations\cite{haug76,elkomoss81}
are directly comparable to the results of this paper, but the 
techniques presented here could be generalized to the other scattering
problems.  Also, the results presented here provide a benchmark for
evaluating the approximations used in other theoretical treatments, 
and could lend insight into the reliability of the approximations 
in more complicated situations. 

One standard theoretical approach is diagrammatic perturbation
theory, as presented in the work of Keldysh and 
Kolsov\cite{keldysh68} and Haug and Hanamura.\cite{haug75}
They estimate the exciton-exciton scattering matrix as arising
from a single term, $\langle k_1+q, k_2-q | H_{\rm int} | k_1, k_2 \rangle$,
where $|k_1,k_2\rangle$ represents a state of two 
noninteracting excitons with momentum $k_1$ and $k_2$ and
$H_{\rm int}$ is the inter-exciton Coulomb interaction.
This method gives an estimate of 
$a_s = \frac{13}{6}a_x$ (independent of the mass ratio),
where $a_x=m_e^{-1}+m_h^{-1}$ is the exciton radius,
but it is an uncontrolled approximation which may have limited
validity in the low energy limit.  One serious drawback of the method 
is that it does not include effects of the biexciton 
in the scattering.  As we show later,
biexciton vibrational states cause strong dependence of
the scattering length on the mass ratio $m_e/m_h$, which is not
captured by the low order perturbation theory.

A second common approach was developed by Elkomoss and 
Munchy,\cite{elkomoss81} and uses an effective 
exciton-exciton potential defined by
$V_{\mathrm eff}(R) = \langle \phi_{\mathrm f}(R) | H
 | \phi_{\mathrm f}(R) \rangle$,
where $\phi_{\mathrm f}(R)$ is the wavefunction for two free
excitons a distance $R$ apart.  The effective potential 
$V_{\rm eff}$ arises from the Hartree term and is
used in a two-particle central-field calculation.
While an exciton-exciton scattering pseudo-potential would be
a very useful tool, this approximate form has some serious drawbacks.
Among its deficiencies are a lack of correlation, 
no van der Waals attraction, a failure to reproduce biexciton
states, and a vanishing interaction potential 
for $m_e = m_h$.  The cross sections calculated by this method are small
and lack qualitative agreement with the results of the present work.

Some insight into exciton-exciton scattering can
be gained by considering the bound states, biexcitons.
Since the number of bound states, $N_{\mathrm B}$, 
enters in the phase shift at zero energy, 
$\delta(0)=\pi N_{\mathrm B}$,
it is necessary that a good computation method for low energy
scattering be able accurately calculate biexciton 
binding energies.\cite{energynote}
For the mass ratios considered (and far beyond, including deuterium)
the biexcitons cannot bind in the  $\phi^{--}$ states,
so biexcitons in rotational
$s$ states always have $\phi^{++}$ wavefunctions.
Detailed theoretical descriptions of biexcitons can be found in Ref. 
[\onlinecite{quattropani77}].
The equal mass case was shown to have a bound biexciton
by Hyllerass and Ore using a variational argument,\cite{hyllerass47}
and a better variational estimate of the binding energy
was given by Brinkman, Rice and Bell,\cite{brinkman73} who found
$E_{\mathrm B} = 0.029 E_x$, where $E_x=0.5/(m_e^{-1}+m_h^{-1})$ 
is the exciton binding energy.
However, because of the importance of correlation
energy, the latter variational treatment was missing {\em half} of the 
biexciton binding energy, as shown by diffusion Monte Carlo (DMC) 
calculations,\cite{lee83,bressanini97}
which find $E_{\mathrm B} = 0.06404(4)\, E_x$.
DMC is a quantum Monte Carlo (QMC) method that uses
a random walk to project out
the ground state wavefunction from a variational wavefunction,
in order to stochastically sample the exact ground state energy.  The success
of DMC for calculating biexciton energies has been a motivation for
its use in the present scattering calculations.  

\section{Method: Quantum Monte Carlo calculation of scattering} 

\subsection{R-Matrix approach and scattering boundary conditions}

The $R$-Matrix approach to scattering is to examine the standing
waves of the system.  As shown by Carlson, Pandharipande,
and Wiringa\cite{carlson84} and Alhassid and Koonin,\cite{alhassid84}
by fixing nodes in the standing waves the scattering problem may
be cast as a stationary state problem suitable for QMC methods.
For an elastic scattering process, we label the distance between
the products by $R$, and the reduced mass of the products by
$m_r$.  In exciton-exciton scattering there is a 
subtlety in the definition of R due to inter-exciton 
exchange, which we will address below in our discussion
of the exciton-exciton scattering wavefunctions.
Nonetheless,
for large separation $R$, the relative motion of the products is
free-particle like, so the many-body wavefunction
depends on $R$ as
\begin{equation}
\phi \propto \sin[k R - \frac{1}{2}l\pi + \delta_l(k)],
\label{asymptoticform}
\end{equation}
where $l$ is the relative angular momentum, $k$ is the scattering
momentum, and $\delta_l(k)$ is the phase shift.
If we constrain the wavefunction to have a node at a large
exciton separation $R_n$, we find
a discrete energy spectrum $E_\alpha(R_n)$, which may be computed by
ground state or excited state methods, such as DMC.
Each choice of $R_n$ gives a spectrum of states ${\phi_a},
\alpha=1,2,\ldots$, with energies ${E_\alpha}$ that determine
values of $\delta_l(k)$,
\begin{equation}
\delta_{l}(k_\alpha)=-k_\alpha R_n +\frac{1}{2}l\pi+\alpha\pi,
\label{kfromE}
\end{equation}
where $k_\alpha=\sqrt{2 m_r E_\alpha}$.  
The scattering matrix elements are determined by the phase 
shifts,
\begin{equation}
S_l(k) = \frac{1}{2ik}\exp[2i\delta_l(k)]-1.
\end{equation}

Carlson has also proposed fixing the logarithmic derivative 
of the wavefunction at the boundary instead of setting wavefunction to zero,
\cite{carlson98}
\begin{equation}
\left.\frac{\hat S \cdot \nabla_{R}\phi}{\phi}\right|_{R=R_b}=\beta,
\end{equation}
were $\hat S$ is the normal to the boundary surface, at a fixed radius $R_b$,
and $\beta$ parameterizes the boundary condition.  This formulation
has the advantage of separating the choice of simulation size $R_b$ 
(subject to $R_b$ lying in the asymptotic region),
from the sampling of energy, which is handled by varying $\beta$,
and is particularly well suited for finding the scattering length.
The application to VMC calculations is straight-forward, but 
preserving the boundary condition in DMC calculations requires 
a method of images.
\cite{carlson99p}
The results presented  here do not use the logarithmic-derivative 
boundary condition.

\subsection{Calculation of excited states}
The use of excited states $\phi_\alpha (\alpha>1)$ is necessary when 
there is a bound state and, more generally, when the scattering state
being studied has its first node before the asymptotic region
is reached.  We use a QMC method to calculate excited states
developed by Ceperley and Bernu,\cite{ceperley88} to
adapt VMC and DMC methods for a Hilbert space of several
low energy wavefunctions.  A set of $m$ trial wavefunctions 
is chosen, $f_\alpha, \alpha=1,\ldots,m$.  The generalized eigenvalue
equation to be solved is,\cite{ceperley88} 
\begin{equation}
\sum\limits_{\beta=1}^{m} [H_{\alpha\beta}(t)-\Lambda_k(t)
N_{\alpha\beta}]d_{k\beta}(t)=0,
\label{eq:eigen}
\end{equation}
where $d_k(t)$ is the $k\mathrm{th}$ eigenvector with eigenvalue
$\Lambda_k(t)$ and the matrices $N$ and $H$ are the 
overlap and Hamiltonian matrices in our trial basis, given by
\begin{eqnarray}
\label{eq:overlap}
N_{\alpha\beta}(t) &=& \int dR_1 dR_2 f^*_\alpha(R_2)e^{-tH}
f^{\phantom *}_\beta(R_1), \\
\label{eq:hmat}
H_{\alpha\beta}(t) &=& \int dR_1 dR_2 H f^*_\alpha(R_2)e^{-tH}
f^{\phantom *}_\beta(R_1).
\end{eqnarray}
The parameter $t$ is the projection time.  
The eigenvalues $\Lambda_k(t)$ are energy eigenvalues
$E_k$ within the Hilbert space spanned by the projected trial functions
$\{e^{-tH/2}f_\alpha\}$, and approach the exact energy eigenvalues
in the limit of large $t$.

The matrices $N$ and $H$ are sampled
with random walks, using a guiding function $\psi$
which must be positive everywhere.
The guiding function must have significant overlap
with all basis functions, and should be optimized to decrease the variance
of the sampled matrices.
At each step $i+1$ in the random walk,
the coordinates $R_{i+1}$ of the particles are updated using
\begin{equation}
R_{i+1} = R_i + \tau \lambda \psi^{-1} \nabla \psi(R_i) 
+ (2\tau \lambda)^\frac{1}{2}\chi_i,
\label{scat_step_eq}
\end{equation}
where $\chi_i$ is a normally distributed random variate with zero mean
and unit variance and $\tau$ is the time
step.   In the limit of small $\tau$, Eq.~(\ref{scat_step_eq}) describes
a process for sampling $\psi^2(R)$.

The matrix element of $e^{-tH}$ is estimated by
integrating the local energy 
of the guiding function, 
$E_{L\psi}(R) = \psi^{-1}(R) H \psi(R)$, along the random walk,
\begin{equation}
W_{n,n+k} = e^{-\tau\!\!\sum\limits_{j=n}^{n+k-1}
\frac{1}{2}[E_{L\psi}(R_j) + E_{L\psi}(R_{j+1})]}.
\end{equation}
The estimators for matrices $N$ and $H$ are
\begin{eqnarray}
\label{eq:overlapest}
n_{\alpha\beta}(k\tau) &=& \frac{1}{p} \sum\limits_{i=1}^p
F^*_\alpha(R_i) W_{i,i+k} F^{\phantom *}_\beta(R_{i+k})
\\
\label{eq:hmatest}
h_{\alpha\beta}(k\tau) &=& \frac{1}{p} \sum\limits_{i=1}^p
F^*_\alpha(R_i) W_{i,i+k} F^{\phantom *}_\beta(R_{i+k})E_{L\beta}(R_{i+k}),
\end{eqnarray}
where $F_\alpha=f_\alpha/\psi$
and $E_{L\beta}= f_\beta^{-1}(R) H f_\beta(R)$ are the
local energies of the trial basis states.

\subsection{Form of exciton-exciton scattering wavefunctions}

\begin{figure}[t]
\includegraphics[width=\linewidth]{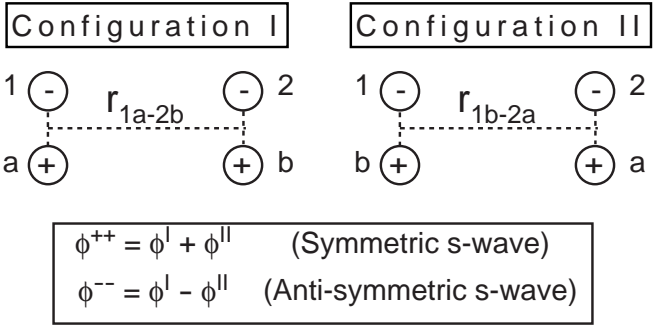}
\caption{The two contributions to the exciton scattering wavefunctions
$\phi$, at large exciton-exciton separation.
The asymptotic form of the $S$-wave scattering states are symmetric
($\phi^{++}$) or antisymmetric ($\phi^{--}$) combinations of these
configurations.  This symmetric/antisymmetric form is used
for the trial wavefunctions $f^{++}$ and $f^{--}$, using
$f^{\text{I}}$ and $f^{\text{II}}$ from Eq.~(\ref{trial_ex_scat_psi_eq}). }
\label{configfig}
\end{figure}

We now discuss the form for the exciton scattering functions
$f^{++}_\alpha$ and $f^{--}_\alpha$.
As mentioned before, inter-exciton
exchange of particles complicates the definition of exciton-exciton
separation.  There are two configurations for well-separated 
excitons, as shown in Fig.~\ref{configfig}.  Configuration I
has the electrons and hole paired as $1a$, $2b$;
and Configuration II as $1b$, $2a$.
We choose wavefunctions $f_\alpha^{\mathrm I}$ and $f_\alpha^{\text{II}}$
to represent these states,
\begin{eqnarray}
f_\alpha^{\text{I}}
   &=& e^{-\gamma r_{1a}}e^{-\gamma r_{2b}}
       U_\alpha(r_{1a-2b})  \nonumber\\
&&e^{\frac{c_{f}r_{12}}{1+d_{f}r_{12}}
     +\frac{c_{g}r_{ab}}{1+d_{g}r_{ab}}
     -\frac{c_{h}r_{1b}}{1+d_{h}r_{1b}}
     -\frac{c_{h}r_{2a}}{1+d_{h}r_{2a}}}, \nonumber\\
f_\alpha^{\text{II}}
   &=& e^{-\gamma r_{1b}}e^{-\gamma r_{2a}}
       U_\alpha(r_{1b-2a}) \nonumber\\
&&e^{\frac{c_{f}r_{12}}{1+d_{f}r_{12}}
     +\frac{c_{g}r_{ab}}{1+d_{g}r_{ab}}
     -\frac{c_{h}r_{1a}}{1+d_{h}r_{1a}}
     -\frac{c_{h}r_{2b}}{1+d_{h}r_{2b}}},
\label{trial_ex_scat_psi_eq}
\end{eqnarray}
where $\gamma, c_f, d_f, c_g, d_g, c_h, d_h$, and parameters 
in the function $U_{\alpha}(r)$ are variational.
These wavefunctions represent two excitons in a relative s-wave state.
Since these are not eigenstates of the exchange operator for
electrons $\hat{P}_e$ or holes, $\hat{P}_h$, we take linear combinations
of the two for our trial wavefunctions, 
$f_\alpha^{++} =\ f_\alpha^{\text{I}} + f_\alpha^{\text{II}}$ and
$f_\alpha^{--} =\ f_\alpha^{\text{I}} - f_\alpha^{\text{II}}$.
For large separation of excitons, the exponential factors
prohibit both configurations from 
simultaneously contributing to the wavefunction.
Thus, a node can be approximated in the scattering wavefunction by 
simply requiring that $U_\alpha(r)$ be zero for all $r>R_n$.  The
error introduced by this approximation is of order $\exp(-2\alpha R_n)$, 
and is another limit on the use of small values for $R_n$.
Since we only do calculations
for low energy scattering, large $R_n$, the lack of a 
well-defined  exciton-exciton
separation distance for short distances does not matter.

\begin{figure}[tb]
\includegraphics[width=\linewidth]{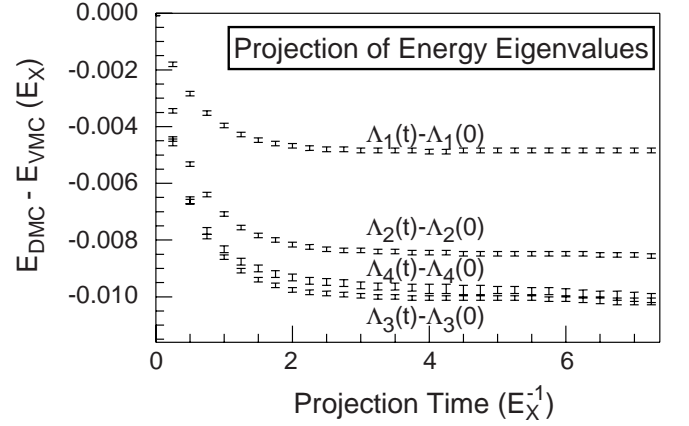}
\caption{Change in the projected energy
eigenvalues $\Lambda_\alpha(t)-\Lambda_\alpha(0)$ of the 
DMC states relative to the VMC states as a function of DMC projection 
time $t$, for basis states $\alpha=1,2,3,4$.  The eigenvalue
equation is given by Eqs.~(\ref{eq:eigen})--(\ref{eq:hmat}), where the
Hamiltonian and overlap matrices have been sampled using
Eqs.~(\ref{eq:overlapest}) and (\ref{eq:hmatest}).
}
\label{EvsTfig}
\end{figure}

This method for calculating scattering properties is 
very sensitive to the energy spectra $\{E_\alpha(R_n)\}$.
To get accurate energies, we do not try to construct and optimize elaborate
variational wavefunctions, but rather use DMC to project the energy 
from trial wavefunctions of the form given in 
Eq.~(\ref{trial_ex_scat_psi_eq}).
The coefficients $\gamma$, $c_f$, $c_g$, and $c_h$ are
chosen to obey the cusp conditions on 
the wavefunction for small particle separations.
The s-wave envelope functions $U_\alpha(r)$
are taken as solutions to an empirical exciton-exciton scattering potentials,
\begin{equation}
V(r)=\left\{
  \begin{array}{ll}
    -V_0\left(1-\frac{3r^2}{4dr^2}\right);\;& r \le d,\\
    -V_0\frac{d^6}{4r^6}                ;\;& r  >  d,
  \end{array}
  \right.
\end{equation}
where $V_0$ and $d$ have been self-consistently fit to
approximate the energy spectrum of the four particle scattering states.
We take the guiding function $\psi$ to have the same form as the
$f^{++}$ wavefunctions with $U_\psi(r)=[ce^{-r^2/8}
+\sum_\alpha d_\alpha U^2_\alpha(r)]^{1/2}$.
The parameters $\{c,d_\alpha\}$ are chosen to bias sampling
towards the collision: $\{c=0,d_1=2,d_2=d_3=d_4=1\}$ for $\phi^{++}$ states
and $\{c=2,d_\alpha=1\}$ for $\phi^{--}$ states
To check for convergence of the energies in DMC, we plot the
energy difference, $E_{\mathrm DMC}-E_{\mathrm VMC}$, as a
function of projection time in Fig.~\ref{EvsTfig}.
We see convergence after a projection time of 3 ${\mathrm E_x^{-1}}$.

\begin{figure}[tb] 
\centerline{\includegraphics[width=\linewidth]{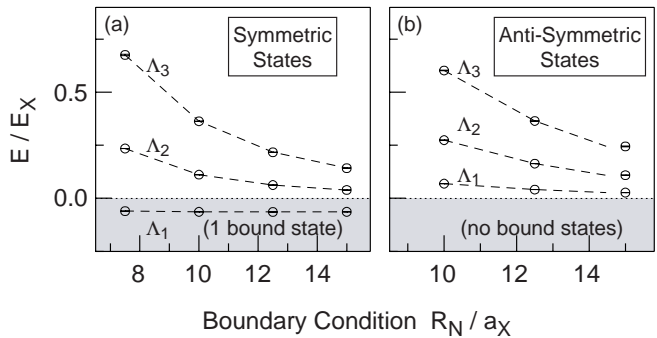}}
\caption{Energy spectra as a function of nodal position $R_N$, for 
(a) symmetric $\phi^{++}$ states, and (b) antisymmetric $\phi^{--}$ states,
with $m_e=m_h$.  The lowest energy curve in (a) is the biexciton with 
binding energy $E_{\mathrm B}=0.0642(3)$.  These functions $E(R_N)$ 
determine the phase shifts $\delta^{++}$ and $\delta^{--}$, 
by the relationship in Eq.~(\ref{kfromE}).}
\label{EvsRnfig}
\end{figure}

We thus find two energy spectra for each value of $R_n$, as shown in 
Figs.~\ref{EvsRnfig}(a) and \ref{EvsRnfig}(b).
The spectra for the symmetric states $\phi^{++}$, show a clearly
bound biexciton state, as seen in Fig.~\ref{EvsRnfig}(a).
The antisymmetric states $\phi^{--}$, as shown in Fig.~\ref{EvsRnfig}(b),
have no bound state.  The binding energy of the biexciton is
$E_{\mathrm B} = -0.0642(3)\,E_x$, in agreement with other
ground state calculations, and is insensitive to the position
of the node $R_N$ because it is localized.
In contrast, the delocalized scattering states
are quite sensitive to $R_N$, and their dependence on $R_N$ is a 
measure of the elastic scattering matrix elements.

\subsection{Calculation of phase shifts}

\begin{figure}[tb]
\centerline{\includegraphics[width=\linewidth]{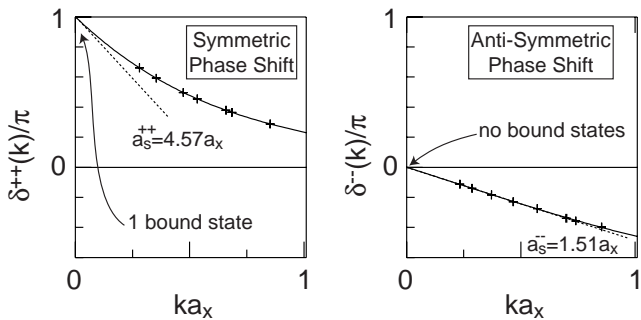}}
\caption{Phase shifts $\delta(k)$ for the two $s$-wave 
scattering states, for $m_e=m_h$, calculate using Eq.~(\ref{kfromE})
and the data from Fig.~\ref{EvsRnfig}.}
\label{phasefig}
\end{figure}

\begin{table}[tb]
\caption{Coefficients for polynomomial fit to the low energy
part of the phase shift functions for the case $m_e=m_h$.}
\begin{ruledtabular}
\begin{tabular}{ccccc}
&$c_0$&$c_1$&$c_2$&$c_3$\\
\hline
$\delta^{++}$&$\pi$&-4.574(21)& 2.995(70)&-0.829(54)\\
$\delta^{--}$&$0$  &-1.512(29)&-0.138(90)& 0.216(64)\\
\end{tabular}
\end{ruledtabular}
\label{phasecoeftab}
\end{table}

Using Eq.~(\ref{kfromE}), we determine scattering phase shifts,
$\delta^{++}(k)$ and $\delta^{--}(k)$, which are shown in 
Fig.~\ref{phasefig} for the equal-mass case.
The $k=0$ limits show us that there is one bound symmetric state 
$\phi^{++}$ and no bound antisymmetric states $\phi^{--}$.
The slope at $k=0$ is related to the scattering length $a_s$ by
$\delta'(k) = -a_s$.
From a cubic polynomial fit to the data,
with coefficients given in Table \ref{phasecoeftab}, we find
$a=4.57(2) a_x$ for the symmetric state, and $a=1.51(3)$ for the 
antisymmetric state.  These model values are consistent with the
measured upper bound $4 a_x$ found in Cu$_2$O.

\subsection{Extracting spin dependence from spatial symmetries}

\begin{table}[tb]
\caption{Matrix elements for changing spin basis in the two exciton problem.
Columns are in the $|s,s_{ee},s_{hh}\rangle_{eh}$ basis and rows are in the 
${}_{ex}\langle s,s_{eh},s_{eh}|$ basis.}
\begin{ruledtabular}
\begin{tabular}{c|cccccc}
&$|000\rangle_{eh}$&$|011\rangle_{eh}$&$|101\rangle_{eh}$&$|110\rangle_{eh}$
&$|111\rangle_{eh}$&$|211\rangle_{eh}$\\
\hline
${}_{ex}\langle000|$&$-\frac{1}{2}$&$\frac{\sqrt3}{2}$& $0$ & $0$ & $0$ & $0$\\
${}_{ex}\langle011|$&$\frac{\sqrt3}{2}$&$+\frac{1}{2}$& $0$ & $0$ & $0$ & $0$\\
${}_{ex}\langle101|$&    $0$    &     $0$   &   $\frac{1}{2}$ & $-\frac{1}{2}$&
  $\frac{\sqrt2}{2}$  & $0$ \\
${}_{ex}\langle110|$&    $0$    &     $0$   &  $-\frac{1}{2}$ &  $\frac{1}{2}$&
  $\frac{\sqrt2}{2}$  & $0$ \\
${}_{ex}\langle111|$&    $0$    &     $0$   &  $\frac{\sqrt2}{2}$&
  $\frac{\sqrt2}{2}$& $0$ & $0$ \\
${}_{ex}\langle211|$& $0$ & $0$ & $0$&  $0$ & $0$ & $1$
\end{tabular}
\end{ruledtabular}
\label{spbasistab}
\end{table}

Single excitons can be in either a spin triplet or spin singlet 
energy eigenstates.  (These states are degerate in our model, 
but are split by electron-hole exchange interaction in more realistic 
models.) During collisions, however, the spin of individual excitons 
is not a conserved quantity --- the excitons
can change their internal spin states during collision. 
To relate our scattering states $\phi^{++}$ and $\phi^{--}$
to excitonic spin states, we invoke the antisymmetry of the 
total (spatial plus spin) fermionic wavefunction.
Fermionic antisymmetry implies that the
states $\phi^{++}$ and $\phi^{--}$ have spin 
eigenstates $|s00\rangle_{eh}$ and $|s11\rangle_{eh}$
where the notation $|s s_e s_h\rangle_{eh}$
stands for total spin $s$, total electron spin $s_e$
and total hole spin $s_e$.
We denote the experimentally relevant ``excitonic spin basis'' as
$|s s_1 s_2\rangle_{ex}$, where 
the total spin is $s$ and the spins on the scattering excitons are
$s_1$ and $s_2$, which take the values  0 for singlet excitons 
and 1 for triplet excitons.
Table~\ref{spbasistab}
lists matrix elements for the change of basis
$|s s_e s_h\rangle_{eh} \rightarrow |s s_1 s_2\rangle_{ex}$.
The states $|0 0 0\rangle_{eh}$ and 
$|s 1 1\rangle_{eh}$ act as channels for s-wave exciton-exciton
scattering.  The matrix elements for scattering through these channels
are:
\begin{eqnarray}
s^{++}(k)&=&{}_{eh}\langle 0 0 0;k|  \hat{s} | 0 0 0;k\rangle_{eh}
=\frac{e^{2i\delta^{++}(k)}-1}{2ik},
\\
s^{--}(k)&=&{}_{eh}\langle S 1 1;k|  \hat{s} | S 1 1;k\rangle_{eh}
=\frac{e^{2i\delta^{--}(k)}-1}{2ik}.
\end{eqnarray}

\begin{table}[tb]
\caption{All non-zero $s$-wave scattering process. Initial and final
states are denoted by $|s s_1 s_2\rangle_{ex}$, where $s$ is the total spin and
$s_1$ and $s_2$ are the individual exciton spins.
The coefficients $\alpha_+,\alpha_-$ are for computing the 
scattering matrix elements, and the coefficients $c_{++},c_{--},c_{+-}$
are for computing cross sections.  The last column list the ratio of the
scattering length $a_s$ to the exciton radius $a_x$ for the case $m_e=m_h$.}
\begin{ruledtabular}
\begin{tabular}{c|rr|rrr|r}
Process&$\alpha_+$&$\alpha_-$&$c_{++}$&$c_{--}$&$c_{+-}$&$a_s/a_x$\\
\hline
$|000\rangle_{ex}\rightarrow|000\rangle_{ex}$
&$\frac{1}{4}$&$\frac{3}{4}$
&$\frac{1}{4}$&$\frac{3}{4}$&$-\frac{3}{16}$&2.128(27)\\
$|011\rangle_{ex}\rightarrow|011\rangle_{ex}$
&$\frac{3}{4}$&$\frac{1}{4}$
&$\frac{3}{4}$&$\frac{1}{4}$&$-\frac{3}{16}$&3.759(22)\\
$|011\rangle_{ex}\rightarrow|000\rangle_{ex}$
&$-\frac{\sqrt3}{4}$&$-\frac{\sqrt3}{4}$
&0&0&$\frac{3}{16}$&-1.411(17)\\
$|000\rangle_{ex}\rightarrow|011\rangle_{ex}$
&$-\frac{\sqrt3}{4}$&$-\frac{\sqrt3}{4}$
&0&0&$\frac{3}{16}$&-1.411(17)\\
$|110\rangle_{ex}\rightarrow|110\rangle_{ex}$
&0&$\frac{1}{4}$&0&1&0&0.706(14)\\
$|211\rangle_{ex}\rightarrow|211\rangle_{ex}$
&0&$1$&0&1&0&1.512(29)
\end{tabular}
\end{ruledtabular}
\label{scattab}
\end{table}

The collision of
two excitons with well defined initial and final spin states can
be determined by decomposing the scattering events into the two channels,
$s^{++}(k)$ and $s^{--}(k)$.  Using the change of basis matrix
(Table~\ref{spbasistab}), we find
$s(k) = \alpha_+ s^{++}(k) + \alpha_- s^{--}(k)$,
where the coefficients $\alpha_+$ and $\alpha_-$ for all non-zero
s-wave scattering processes are given in Table \ref{scattab}.
The s-wave scattering cross sections are given by
$\sigma(k)=8\pi|s(k)|^2$,
where there is a factor of two enhancement due to the identical particle
statistics.  The spin-dependent cross sections take the form,
\begin{eqnarray}
\sigma(k)&=&\frac{8\pi}{k^2}\big[
c_{++}\sin^2\delta^{++} + c_{--}\sin^2\delta^{--}\nonumber\\
&&-c_{+-}\sin^2(\delta^{++}-\delta^{--}) \big],
\label{eq:cs}
\end{eqnarray}
where $c_{++}=\alpha_+(\alpha_++\alpha_-)$,
$c_{--}=\alpha_-(\alpha_++\alpha_-)$,
and $c_{+-}=\alpha_+\alpha_-$
depend on the initial and final spin states and
are tabulated in Table \ref{scattab}.
The scattering lengths $a_s$ are given by $a_s=-\alpha_+\delta^{++'}(0)
-\alpha_-\delta^{--'}(0)$, where the derivatives of the phase shifts
are determined from the linear coefficients in Table \ref{phasecoeftab}.

\begin{figure}[tb] 
\centerline{\includegraphics[width=0.85\linewidth]{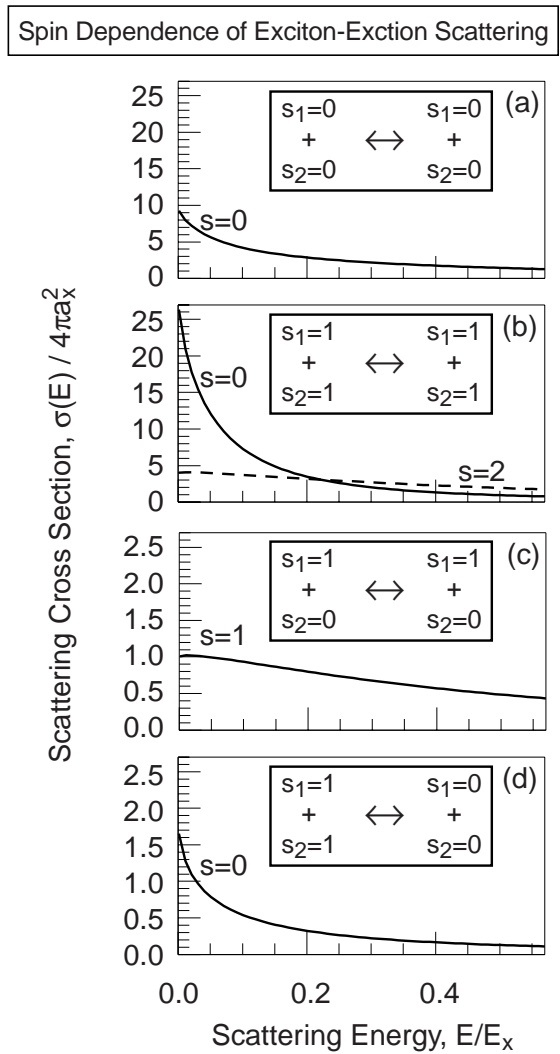}}
\caption{All non-zero s-wave cross sections, Eq.~(\ref{eq:cs}),
for exciton-exciton scattering with $m_e=m_h$, for the processes: (a) singlet-singlet $\rightarrow$ singlet-singlet, (b) triplet-triplet $\rightarrow$ triplet-triplet for total spin $s=0$ (dashed line) and $s=2$ (solid line), (c) triplet-singlet $\rightarrow$ triplet-singlet with $s=1$, and (d) triplet-triplet $\rightarrow$ singlet-singlet and singlet-singlet $\rightarrow$ triplet-triplet, both with $s=1$.}
\label{cs1fig}
\end{figure}

\section{Results}

\subsection{Spin dependence of scattering cross sections for the equal 
mass case}

The calculated spin-dependent scattering lengths for the case $m_e=m_h$ 
are presented in the last column  of Table~\ref{scattab}.
These are the low-energy limit the cross-sections shown in 
Fig.~\ref{cs1fig}. In
Fig.~\ref{cs1fig} we have plotted the s-wave scattering cross sections 
versus scattering momentum for the case $m_e=m_h$ and 
all non-zero spin configurations.
Fig.~\ref{cs1fig}(a) shows scattering of two singlet-excitons.
Scattering of two triplet-excitons is shown in 
Fig.~\ref{cs1fig}(b), where the solid line represents
the spin aligned $S=2$ state, and the dashed line represents
the $S=0$ state.  The $S=0$ state scatters particularly strong
because it has a large contribution from the $s^{++}$ channel,
which is enhanced by the weakly bound biexciton.  Triplet-excitons
in a relative $S=1$ state are spatially antisymmetric and 
thus have no s-wave scattering.
We show s-wave scattering of triplet-excitons from singlet-excitons in 
Fig.~\ref{cs1fig}(c).  This state has two distinguishable excitons,
and can scatter by both s-wave and p-wave processes.  As can be
seen in Table \ref{scattab}, the only contribution to the 
cross section is from 
the weaker $s^{--}$ channel.  The coefficient for s-wave
scattering is particularly small because only half the 
scattering process is symmetric (s-wave), and 
there is an additional factor 
of one-half which cancels the identical particle factor.

There is also an triplet- to singlet-exciton conversion cross section given
in Fig.~\ref{cs1fig}(d).  Although this is an inelastic process
in experimental situations,
it conserves energy according to our model Hamiltonian because
we do not have an explicit interband spin coupling.
The conversion of two triplet-excitons to two singlet-excitons can be 
understood as an inter-exciton 
exchange of a pair of electrons (or holes).  Since the spins of the 
individual excitons do not correspond to symmetries of the Hamiltonian,
they need not remain constant during scattering.  This conversion
process is a physical consequence of the two inequivalent scattering channels
$s^{++}$ and $s^{--}$.  This effect has been reported in 
experimental\cite{amand97} and theoretical\cite{cuiti98} 
work on exciton scattering in quantum wells.

\subsection{Mass dependence of scattering lengths}

\begin{figure}[tb]
\includegraphics[width=\linewidth]{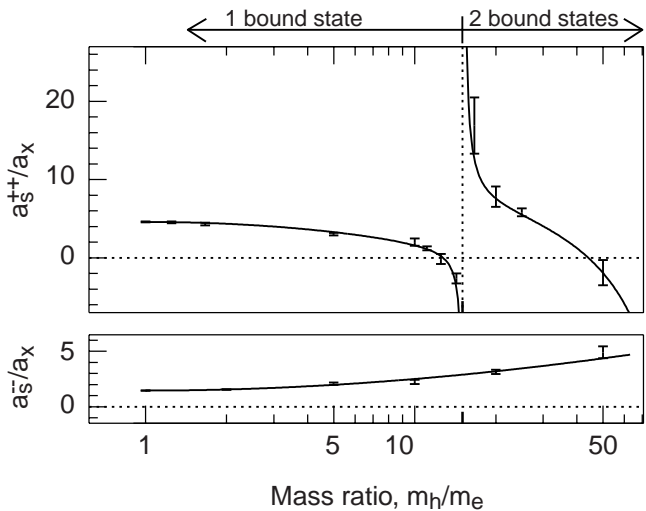}
\caption{The scattering lengths $a_s^{++}$ and $a_s^{--}$ as a function 
of the electron-hole mass ratio.  The divergence in $a_s^{++}$ near 
$m_h/m_e = 15$ is due to the appearance of second bound biexciton
state.  Solid lines are a guide to the eye.}
\label{avmfig}
\end{figure}

The dependence of the cross sections on mass may be numerically
studied by our methods.  In Fig.~\ref{avmfig} we show our calculated
$a_s^{++}$ and $a_s^{--}$ as a function of mass ratio, $m_h/m_e$.
We find that the scattering length is remarkably insensitive to
the mass ratio for $m_e/m_h < 10$ (corresponding to a wide range 
of semiconductors), but then diverges near $m_e/m_h = 15$.
This feature is lost in previously published theoretical 
treatments of exciton-exciton scattering.
The divergence in $a_s^{++}$ near $m_h/m_e = 15$ is due to the acquisition
of a biexciton vibrational state.  For $m_e=m_h$ the biexciton has no
bound excited states, while a $\mathrm H_2$ molecule has 15 bound 
vibrational states.  Our calculations have shown the first of these
appears near $m_h/m_e = 15$, with dramatic effects on the scattering
length.  The $a_s^{--}$ curve is relatively featureless because there
are are no bound antisymmetric states in this range.
We interpret the upward drift of $a_s^{--}$ for larger mass ratios 
as a systematic error due to difficulties in projecting 
states in excitonic systems with very different electron and hole
masses.  The heavy particle determines the projection time
$(\sim m^{-1}R^{-2})$ while the light particle determines the 
diffusion time step $\tau$.  The difficulty in handling large mass
ratios make the method (as presented here)
complementary to calculations that use the Born-Oppenheimer approximation.

It is important to realize that similar relationships must
exist between the scattering length and
other material parameters, such as the Luttinger-Kohn parameters
describing realistic hole states, external strain, and spin-orbit
coupling, to name a few.  Theoretical studies of such effects will
need similar high-accuracy scattering calculations, but applied to
more accurate Hamiltonians, and are an area for future research.

\section{Conclusion}

To summarize, we have shown that there are several significant
elastic scattering processes for excitons, and have given numerically
exact values for a widely used theoretical model.
We find strong triplet-triplet and  singlet-singlet scattering, with weaker
triplet-singlet scattering and triplet-triplet to 
singlet-singlet conversion processes.
Scattering is relatively insensitive to the mass ratio 
for $m_e/m_h \alt 10$, but becomes very sensitive and actually
diverges near $m_e/m_h \approx 15$.
DMC has been found be a good tool for this 
four-particle excited state calculation, since the
detection of weakly bound states requires very accurate
evaluation of the correlation energy.

This computational approach should be extended in many ways.
The extension to higher angular momentum states would
give important corrections at higher scattering energies.
Application to biexciton-biexciton scattering are possible, but
would be a bit more difficult because the scattering wavefunction would
then have to describe eight interacting particles.
Most importantly, the method should be adapted to better Hamiltonians
so that the sensitivity of the scattering length to material properties
for a wide range of materials can be studied.  This approach
should be quite useful for quantum well problems, which have
similarities to the bulk problem study here, but have many more
experimental parameters that can affect exciton-exciton interaction.

\acknowledgments
We would like to thank K.~O'Hara and J.~Carlson for useful discussions.
This work was supported by NSF Grant No. DMR 98-02373,
computer resources at NCSA, and the Department of Physics at
the University of Illinois Urbana-Champaign.


\begin{thebibliography}{10}
\bibitem{energynote}
For example, the location of the scattering length divergence
at $m_e/m_h\approx 15$, reported later in this article, is
extremely sensitive to an accurate treatment of correlation energy,
since it relies on the appearance of a second, weakly 
bound biexciton state at that mass ratio.
\end{thebibliography}
\end{document}